\documentclass[12pt]{article}
\usepackage{amsmath}
\usepackage{amssymb}
\usepackage{latexsym}
\usepackage{url}

\usepackage[all]{xy}

\newcommand{\edge}{
\bsegment 
  \rlvec (0 10) \onedot 
  \savepos (0 10)(*ex *ey)
\esegment
\move (*ex *ey)
}

\newcommand{\edgesV}{
\bsegment
  \move (0 0)
  \rlvec (-4 10) \onedot
  \move (0 0)
  \rlvec (4 10) \onedot
  \savepos (4 10)(*ex *ey)
\esegment
\move (*ex *ey)
}

\newcommand{\ctreeone}{
\begin{texdraw}
  \move (0 0) \onedot
\end{texdraw}
}

\newcommand{\ctreetwo}{
\begin{texdraw}
  \move (0 0) \onedot
  \edge
\end{texdraw}
}

\newcommand{\ctreethreeV}{
\begin{texdraw}
  \move (0 0) \onedot
  \edgesV
\end{texdraw}
}

\newcommand{\ctreethreeL}{
\begin{texdraw}
  \move (0 0) \onedot
  \edge \edge
\end{texdraw}
}

\newcommand{\ctreefourL}{
\begin{texdraw}
  \move (0 0) \onedot
  \edge \edge \edge
\end{texdraw}
}

\newcommand{\ctreefourY}{
\begin{texdraw}
  \move (0 0) \onedot
  \edge \edgesV
\end{texdraw}
}

\newcommand{\ctreefourVL}{
\begin{texdraw}
  \move (0 0) \onedot
  \edgesV \move (-4 10) \edge
\end{texdraw}
}

\newcommand{\smalldot}{
  \bsegment
    \move (0 0) \fcir f:0 r:1.5
  \esegment
}

\newcommand{\tinydot}{
  \bsegment
    \move (0 0) \fcir f:0 r:1.1
  \esegment
}

\newcommand{\twotree}{
\bsegment 
\move (0 -4) \lvec (0 0) \tinydot
\lvec (-2 5) \tinydot
\lvec (-4 10)
\move (-2 5) \lvec (0 10)
\move (0 0) \lvec (3 10)
\esegment
}

\newcommand{\onetree}{
\bsegment 
\move (0 -4) \lvec (0 0) \tinydot
\lvec (-2 5) 
\move (0 0) \lvec (2 5)
\esegment
}

\newcommand{\zerotree}{
\bsegment 
\move (0 -4) \lvec (0 4)
\esegment
}

\newcommand{\twonodetree}{
\bsegment 
\move (0 -4) \smalldot \lvec (0 4) \smalldot
\esegment
}

\newcommand{\vtree}{
\bsegment 
\move (0 -4) \smalldot \lvec (-3 4) \smalldot
\move (0 -4) \lvec (3 4) \smalldot
\esegment
}

\usepackage[dvipsnames]{color}
\newcommand{\red}[1]{{\textcolor{red}{#1}}}
\newcommand{\blue}[1]{{\textcolor{blue}{#1}}}
\newcommand{\darkgreen}[1]{{\textcolor[rgb]{0,0.6,0}{#1}}}
\newcommand{\magenta}[1]{{\textcolor{magenta}{#1}}}

\usepackage{theorem}
\theorembodyfont{\normalfont\itshape}
\theoremstyle{change}
\newtheorem{lemma}{Lemma.}[section]

\newtheorem{theorem}[lemma]{Theorem.}

\theorembodyfont{\normalfont}
\newtheorem{rmk}[lemma]{Remark.}
\newtheorem{taller}[lemma]{$\!\!$}
\newenvironment{blanko}[1]%
{\begin{taller}{\normalfont\bfseries #1}\normalfont}%
{\end{taller}}

\providecommand{\qed}{\hspace*{\fill}$\Box$}

\newenvironment{proof}{%
\begin{list}{\em Proof. }%
{\setlength{\labelsep}{0mm}\setlength{\leftmargin}{0mm}%
\setlength{\labelwidth}{0mm}\setlength{\listparindent}{\parindent}%
\setlength{\parsep}{\parskip}\setlength{\partopsep}{0mm}}%
\item}{\qed\end{list}}

\newenvironment{proof*}[1]{%
\begin{list}{\em #1 }%
{\setlength{\labelsep}{0mm}\setlength{\leftmargin}{0mm}%
\setlength{\labelwidth}{0mm}\setlength{\listparindent}{\parindent}%
\setlength{\parsep}{\parskip}\setlength{\partopsep}{0mm}}%
\item}{\qed\end{list}}

\newcounter{dummycounter}

\newenvironment{punkt-i}%
{%
	\begin{list}%
	{(\roman{dummycounter})}%
	{\usecounter{dummycounter}%
	\setlength{\itemsep}{0em}\setlength{\parsep}{0em}\setlength{\topsep}{0em}%
	\setlength{\itemindent}{0em}\setlength{\labelwidth}{1.8em}%
	\setlength{\labelsep}{0.6em}\setlength{\leftmargin}{2.4em}}%
}%
{\end{list}}

\newenvironment{acknowledgments}{%
\begin{list}{\textbf{Acknowledgments. }}%
{\setlength{\labelsep}{0mm}\setlength{\leftmargin}{0mm}%
\setlength{\labelwidth}{0mm}\setlength{\listparindent}{\parindent}%
\setlength{\parsep}{\parskip}\setlength{\partopsep}{0mm}}%
\item}{\end{list}}

\providecommand{\norm}[1]{\left| {#1}\right|}

\usepackage{mathrsfs}

\newcommand{\N}{\mathbb{N}}
\newcommand{\HH}{\mathscr{H}}
\newcommand{\CK}{\mathscr{H}_{\mathrm{CK}}{}}
\newcommand{\BB}{\mathscr{B}}
\newcommand{\C}{\mathscr{C}}
\newcommand{\tensor}{\otimes}

\newcommand{\Id}{\operatorname{Id}}

\newcommand{\Aut}{\operatorname{Aut}}

\newcommand{\Set}{\mathbf{Set}}
\newcommand{\alg}{\textbf{-}\mathbf{alg}}
\newcommand{\Type}{\mathrm{Type}}
\newcommand{\asbefore}{\texttt{"}}

\newcommand{\isopilback}{\stackrel{\raisebox{0.1ex}[0ex][0ex]{\(\sim\)}}%
			{\raisebox{-0.15ex}[0.28ex]{\(\leftarrow\)}}}

\usepackage{texdraw}
\input{txdtools}

\everytexdraw{%
	\drawdim pt \linewd 0.5 \textref h:C v:C
	\setunitscale 1
}

\newcommand{\onedot}{
  \bsegment
    \move (0 0) \fcir f:0 r:2
  \esegment
}

\begin{document}

\title{Combinatorial Dyson--Schwinger equations and inductive data types}

\author{
Joachim Kock
\thanks{Email address: \texttt{kock@mat.uab.cat}}
\\
{\footnotesize Departament de matem\`atiques}\\[-5pt]
{\footnotesize Universitat Aut\`onoma de Barcelona}
}

\date{}

\maketitle

\begin{abstract}
  The goal of this contribution is to explain the analogy between
  combinatorial Dyson--Schwinger equations and inductive data types
  to a readership of mathematical physicists.  The connection relies
  on an interpretation of combinatorial Dyson--Schwinger equations
  as fixpoint equations for polynomial functors (established
  elsewhere by the author, and summarised here), combined with
  the now-classical fact that polynomial functors provide semantics
  for inductive types.  The paper is expository, and comprises also
  a brief introduction to type theory.
\end{abstract}

%
%

\section*{Introduction}

The aim of this contribution is to point out and explain some connections between
Dyson--Schwinger equations, as employed in quantum field theory (QFT), and inductive data
types as they appear in constructive type theory, at the foundations of
mathematics.  The paper is expository and targets a readership of mathematical
physicists with no background in category theory or type theory, attempting to
explain the required background knowledge along the way.

Briefly, the combinatorial Dyson--Schwinger equations are regarded as syntactic
specifications of inductive types: the $B_+$-operators play the role of
constructors, whereas the equations themselves, which are manifestly fixpoint equations, express
the inductive character of the type, which in type theory is given by
eliminators.  While this connection is, to some extent, a banality --- the inductive
character of Feynman graphs is plain, and the only access we have to the
solutions of the Dyson--Schwinger equations is through induction in some form or
another --- nevertheless, I believe it is  potentially useful to formalise this
analogy and to develop it further.  I would like to suggest that the `natively
inductive' methods of type theory could, perhaps, be useful for structuring
computations in quantum field theory.  On the other hand, there are several
algebraic structures related to Dyson--Schwinger equations, such as Hopf
algebras and pre-Lie algebras, which could possibly be of interest also in type
theory.  I should admit from the outset that I am relatively ignorant of the finer
details of both quantum field theory and type theory, and that the potential
consequences of the presented connections are speculative at present.  My path into
these questions originated in experience with polynomial functors, which are, 
in my view, an important vehicle for establishing the connection.

The polynomial-functor approach to combinatorial Dyson--Schwinger equations is
developed in detail elsewhere \cite{Kock:Poly-DSE}.  Its main point, explained
in the first part of this paper, is to lift the equations from the
Hopf-algebraic level to the objective combinatorial level, dealing directly with
the combinatorial objects themselves through explicit bijections and
categorical methods.  The second part interprets the same ideas in type theory,
exploiting a well-known fact in the categorical semantics of type theory
\cite{Moerdijk-Palmgren:Wellfounded}, namely that least fixpoints for polynomial
functors correspond precisely to W-types, a certain class of inductive types.

%
%
%


\section{Combinatorial Dyson--Schwinger equations}

The Dyson--Schwinger equations are the quantum equations of motion, and are
comprised of infinite hierarchies of functional equations. Solving these equations is a significant
challenge, especially in the field of quantum chromodynamics, where perturbative methods are 
impossible below the confinement scale~\cite{Roberts:1203.5341}.
While there is, of course, an essential analytic aspect to the Dyson--Schwinger
equations, which involves Feynman integrals, there is also a structural aspect that
is essentially related to combinatorics.  For a long period of time, this combinatorial
aspect was characteristic for {\em perturbative} QFT, beginning with Feynman, progressing via
Bogoliubov, Parasiuk, Hepp, and Zimmerman, and culminating in the work of
Kreimer~\cite{Kreimer:9707029} and his collaborators around the turn of the millennium, when this
combinatorics was distilled into clear-cut algebraic structures with numerous
connections to many fields of mathematics~\cite{Connes-Kreimer:9808042}. 
In particular,
Kreimer~\cite{Kreimer:9707029} discovered that the combinatorics of perturbative
renormalisation is encoded in a Hopf algebra of trees, now called the
Connes--Kreimer Hopf algebra. However, it is gradually becoming clear 
\cite{Bergbauer-Kreimer:0506190}, \cite{Kreimer:0509135} that this
combinatorial and algebraic insight is also valuable in
the {\em non-perturbative} regime.

The solution of the full Dyson--Schwinger equations can be expressed in the
form of an infinite sum of integrals. In that case, the solution to
the so-called {\em combinatorial} Dyson--Schwinger equations, introduced by
Bergbauer and Kreimer~\cite{Bergbauer-Kreimer:0506190} and recalled below, requires
determination of this sum's index.  The remaining task essentially involves application of
the Feynman rules.  
The combinatorial Dyson--Schwinger equations are
formulated inside a preexisting combinatorial Hopf algebra, typically the
Connes--Kreimer Hopf algebra, which we begin by reviewing.  While these Hopf
algebras of graphs or trees belong to the perturbative regime, they contain
smaller Hopf algebras spanned by the solutions to the combinatorial
Dyson--Schwinger equations, which also have a non-perturbative meaning.

\begin{blanko}{The Connes--Kreimer Hopf algebra of (rooted) trees.}
  For discussions of the notions of bialgebra and Hopf algebra, see the contribution of 
  Weinzierl~\cite{Weinzierl:1506.09119} in the present volume.
  The {\em Connes--Kreimer Hopf algebra} of (rooted) trees (also called the
  {\em Butcher--Connes--Kreimer Hopf algebra})
  is the free 
  algebra $\CK$ on the set of isomorphism classes of
  {\em combinatorial trees}, such as \
  \raisebox{1pt}{\begin{texdraw}\smalldot\end{texdraw}} , \raisebox{-3pt}{
    \begin{texdraw}\twonodetree\end{texdraw}} , \raisebox{-3pt}{
    \begin{texdraw}\vtree\end{texdraw}} .
    (`Combinatorial' as opposed to the operadic trees (\ref{operadictrees})
    that will play an important role in what follows.)
  
  The comultiplication is
  given on generators by
  \begin{eqnarray*}
  \Delta:  \CK & \longrightarrow & \CK \otimes \CK  \\
    T & \longmapsto & \sum_c P_c \otimes S_c ,
  \end{eqnarray*}
  where the sum is over all admissible cuts of $T$; the left-hand factor $P_c$
  is the forest (interpreted as a monomial) found above the cut, and $S_c$ is
  the subtree found below the cut (or the empty forest, in the case where the cut is below
  the root).  Admissible cut means: either a subtree containing the root, or the empty set.
  $\CK$ is a connected bialgebra: the grading is by the number of nodes, and
  $(\CK)_0$ is spanned by the unit.  Therefore, by general principles (see for 
  example \cite{Figueroa-GraciaBondia:0408145}), it acquires
  an antipode and becomes a Hopf algebra.
%
\end{blanko}

\begin{blanko}{Combinatorial Dyson--Schwinger equations.}\label{BK-DSE}
  The {\em combinatorial Dyson--Schwinger equations} of Bergbauer and
  Kreimer~\cite{Bergbauer-Kreimer:0506190} refer to an ambient combinatorial
  Hopf algebra $\HH$ and a collection of Hochschild $1$-cocycles.  By {\em
  Hochschild $1$-cocycle} is meant a linear operator $B_+$ satisfying the
  equation
  $$
  \Delta \circ B_+ = \big((\Id\tensor B_+) + (B_+ \!\tensor \eta\varepsilon)\big) \circ \Delta , 
  $$
  where $\varepsilon$ is the counit and $\eta$ is the algebra unit.
  The general form of the Dyson--Schwinger
  equation considered by Bergbauer and Kreimer~\cite{Bergbauer-Kreimer:0506190} 
  is
  \begin{equation}\label{eq:BK-DSE}
    X = 1 + \sum_{n\geq 1} w_n \alpha^n \; B^n_+ (X^{n+1})  .
  \end{equation}
  Here $B^n_+$ is a sequence of $1$-cocycles, $w_n$ are scalars, and the
  parameter $\alpha$ is a coupling constant.  The solution $X$ is a formal
  series, an element in $\HH[[\alpha]]$.  By making the ansatz $X= \sum_{k\geq
  0} c_k \alpha^k$, substituting it into the equation, and solving for powers of
  $\alpha$, it is easy to see that there exists a unique solution, which can be
  calculated explicitly up to any given order, as exemplified below.
\end{blanko}

\begin{theorem}[Bergbauer and Kreimer~\cite{Bergbauer-Kreimer:0506190}]
  The $c_k$ span a Hopf subalgebra of $\HH$, which is isomorphic to the Fa\`a di Bruno
  Hopf algebra.
\end{theorem}
The importance of this result is that while $\HH$ is inherently of perturbative
nature, the Fa\`a di Bruno Hopf subalgebra spanned by the solution has a meaning
also non-perturbatively.  In practice, $\HH$ is a Hopf algebra of Feynman graphs,
but for many purposes one can reduce to $\CK$, the Hopf algebra of combinatorial
trees.  In this case, there is only one $B_+$-operator, namely the one that
receives as input a forest and grafts the trees in onto a new root node to
produce a single tree.  The following three examples refer to this Hopf algebra.

\begin{blanko}{Example: A linear Dyson--Schwinger equation.}\label{ex:linear-DSE}
  Consider the equation
  $$
  X=1+\alpha B_+(X) ,
  $$  
  but note that the exponent of $\alpha$ does not
  adhere to the general form in \eqref{eq:BK-DSE}.  With $X = \sum_{k\geq 0} 
  c_k \alpha^k$, one readily finds
  $$
  c_0 = 1, \quad c_1 = \raisebox{1pt}{\ctreeone}, \quad c_2 = \raisebox{-2pt}{\ctreetwo} , 
  \quad c_3 = 
  \raisebox{-4pt}{\ctreethreeL}, \quad c_4 = 
  \raisebox{-8pt}{\ctreefourL},  \quad \text{ etc.}
  $$
  This is the ladder case; although it is of relevance in
  QFT~\cite{Mencattini-Kreimer:0408053}, our interest here is mainly that it is,
  in a precise sense, an expression of recursion in its purest form, as we shall
  see in Section~\ref{sec:N}.
\end{blanko}

\begin{blanko}{Example: A quadratic Dyson--Schwinger 
  equation~\cite{Bergbauer-Kreimer:0506190}.}\label{ex:binary-DSE}
  In the example
  $$
  X=1+\alpha B_+(X^2) ,
  $$
  substituting the ansatz $X = \sum_{k\geq 0} c_k \alpha^k$
  into the equation and solving the result for powers of $\alpha$
  readily yields
  $$
  c_0 = 1, 
  \ c_1 = \raisebox{1pt}{\ctreeone} , 
  \ c_2 = 2 \raisebox{-2pt}{\ctreetwo} ,
  \ c_3 = 4 \raisebox{-4pt}{\ctreethreeL} + \raisebox{-2pt}{\ctreethreeV} ,
  \ c_4 = 8 \raisebox{-8pt}{\ctreefourL} 
  + 2 \!\!\raisebox{-4pt}{\ctreefourY} + 4 \raisebox{-4pt}{\ctreefourVL} ,
  \quad \text{etc.}
  $$
\end{blanko}

\begin{blanko}{Example: More complex trees.}\label{ex:infinite-DSE}
  We finally consider the following `infinite' example:
  $$
  X = 1 + \sum_{n\geq 1} \alpha^n \; B_+ (X^{n+1}) .
  $$
  With $X= \sum_{k\geq 0} c_k \alpha^k$ again, one finds 
  $$
  c_0 = 1, \ \
  c_1 = \raisebox{1pt}{\ctreeone} , \ \
  c_2 = 2 \raisebox{-2pt}{\ctreetwo} + \raisebox{1pt}{\ctreeone}, \ \
  c_3 = 4 \raisebox{-4pt}{\ctreethreeL} + \raisebox{-2pt}{\ctreethreeV}
  + 5 \raisebox{-2pt}{\ctreetwo} + \raisebox{1pt}{\ctreeone},
\phantom{xxxxxxxxxxxxxxx}
  $$
  $$
\phantom{xxxxxxxxxxxx}
  \ \ c_4 = 8 \raisebox{-8pt}{\ctreefourL} 
  + 2 \raisebox{-4pt}{\ctreefourY} + 4 \raisebox{-4pt}{\ctreefourVL} 
  + 16 \raisebox{-4pt}{\ctreethreeL} + 5 \raisebox{-2pt}{\ctreethreeV}
  + 9 \raisebox{-2pt}{\ctreetwo} + \raisebox{1pt}{\ctreeone},
  \quad \text{etc.}
  $$
  The pattern here may not be obvious.  The actual mechanism
  will only become clear once we move to the setting of polynomial functors
  (see in particular \ref{core}).
\end{blanko}

\section{Polynomial functors and initial algebras}
\label{sec:poly}

While the Dyson--Schwinger equations are traditionally formulated inside a
preexisting combinatorial Hopf algebra, typically in the style of Connes--Kreimer, the
following abstraction steps were taken in \cite{Kock:Poly-DSE}. One begins with
an abstract polynomial fixpoint equation formulated in sets (or groupoids),
without reference to Hopf algebras or $B_+$-operators.  The solution is always
a set (or groupoid) of certain operadic trees, and these trees 
automatically form a Connes--Kreimer-like bialgebra.  Inside this bialgebra,
there are canonical $B_+$-operators (although they are {\em not} Hochschild
$1$-cocycles), in terms of which the original equation can be internalised
to the bialgebra.  The solution, a groupoid of trees, always spans a
sub-bialgebra isomorphic to the Fa\`a di Bruno bialgebra.  In this way, each
equation defines its own bialgebra, a canonical home for it.  However, these
bialgebras are related: every cartesian natural transformation between
polynomial functors yields a homomorphism between the associated bialgebras,
along which the solutions to the Dyson--Schwinger equations are preserved.  A
special case of this is the cartesian subfunctors: these correspond to
truncations which automatically produce sub-bialgebras.  Finally, every such
bialgebra of $P$-trees comes with a canonical bialgebra homomorphism to the
Connes--Kreimer Hopf algebra, which therefore also receives a plethora of
different Fa\`a di Bruno sub-bialgebras.

In this section, we briefly elaborate on thses mathematical constructs, 
referring the reader to \cite{Kock:Poly-DSE} for all details.




\begin{blanko}{Categories and functors.}
  The setting for the material in this section is category theory, but very little is
  required for the level of detail presented here.  For further background
  information on category theory,
  see Leinster~\cite{Leinster:BCT} for a short and concise introduction,
  and Spivak~\cite{Spivak:CTS} for an account targeted at non-mathematicians.
  
  A {\em category} has objects and morphisms, and morphisms can be composed.
  A basic example is the category of sets, denoted $\Set$, where the objects
  are sets and the morphisms are maps of sets.  Other examples are the category
  $\mathbf{Vect}$ of vector spaces and linear maps, or the category
  $\mathbf{Bialg}$ of bialgebras and bialgebra homomorphisms.  A {\em functor}
  is a `morphism of categories,' i.e., it sends objects to objects and morphisms to
  morphisms, in such a way as to preserve composition.  For example, there is a
  functor $F: \Set \to \mathbf{Vect}$ sending a set $S$ to the vector space
  spanned by $S$ and sending a set map $f:S \to T$ to the
  linear map induced by its value on the basis vectors.
\end{blanko}

We work in the category $\Set$ of sets and set maps.

\begin{blanko}{Polynomial functors.}
  The theory of polynomial functors has roots in many different fields of 
  mathematics and computer science, but the work of unifying these developments
  is more recent~\cite{Gambino-Kock:0906.4931}.  The basic notion required here
  are elementary:
  
  Given a map of sets $p: E\to B$, we define a {\em polynomial functor} as
\begin{eqnarray}
  P:\Set & \longrightarrow & \Set \notag  \\
  X & \longmapsto & \sum_{b\in B} X ^{E_b} .\label{eq:poly}
\end{eqnarray}
In the formula, the sum sign denotes disjoint union of sets, and $E_b =
p^{-1}(b)$ denotes the inverse image of an element $b\in B$.  The exponential
notation $X^A$ stands for the set of maps from $A$ to $X$; this notation
(standard in category theory) is justified by the fact that if $X$ is an
$n$-element set and $A$ is a $k$-element set, then $X^A$ is an $n^k$-element
set.  We see that the role played by the map $p:E \to B$ is to deliver a
family of sets indexed by $B$, namely
\begin{equation}
\label{eq:setfamily}
( E_b \mid b\in B).
\end{equation}
(Note that $B$ may be an infinite set, in which case the sum in \eqref{eq:poly} is, accordingly, 
 infinite. Therefore, in a sense, polynomial functors are more
like power series than polynomials.)

To say that $P$ is a {\em functor} means that it operates not just on sets, but 
also on maps: given a map of sets $a:X \to Y$, there is induced a map
$$
\sum_{b\in B} X ^{E_b} \to \sum_{b\in B} Y ^{E_b}
$$
termwise given by
\begin{eqnarray*}
  X^{E_b} & \longrightarrow & Y^{E_b}  \\
  f & \longmapsto & a\circ f.
\end{eqnarray*}
\end{blanko}

\begin{blanko}{Polynomial fixpoint equations.}
  The abstract combinatorial `Dyson--Schwinger equations' we consider here
  are equations of the form
  \begin{equation}\label{eq:polyfix}
  X \isopilback 1 +P(X) ,
  \end{equation}
  where $P$ is a polynomial functor and
  $1$ denotes a singleton set.  This is an equation of sets, and to solve
  this equation means to find a set $X$ together with a specific bijection with $1+P(X)$,
  as indicated by the symbol $\isopilback$.  In fact, we are not satisfied with
  finding just {\em any} solution, rather, we require the {\em best} solution,
  the {\em least fixpoint}.  Making this notion precise requires some further 
  concepts from category theory: 
\end{blanko}
  
\begin{blanko}{Initial objects.}
  An object $I$ in a category $\C$ is called {\em initial} if for every object 
  $C$ there is a unique morphism in $\C$ from $I$ to $C$.  (See the Appendix for 
  further discussion.)  It is easy to show that an initial object, if it 
  exists, is unique (up to isomorphism).  
  For example, the category of sets
  has an initial object, namely the empty set $\emptyset$:
  for any set $X$ there is a unique
  set map $\emptyset \to X$.  (As another example, the category of rings has
  an initial object, namely the ring $\mathbb{Z}$.)
\end{blanko}

\begin{blanko}{$P$-algebras and initial algebras.}
  A {\em $P$-algebra} is by definition a pair $(A,a)$ where $A$ is a set and
  $a: P(A) \to A$ is a set map.  A {\em homomorphism} of $P$-algebras from $(A,a)$ to 
  $(B,b)$ is a set map $f:A\to B$ compatible with the structure maps $a$ and $b$,
  i.e.~such that this square commutes:
  $$\xymatrix{
     P(A) \ar[r]^-a\ar[d]_{P(f)} & A \ar[d]^f \\
     P(B) \ar[r]_-b & B .
  }$$
  Note that the functoriality of $P$ is necessary even for this
  compatibility to be stated: we require the ability to evaluate $P$ not just on sets, but also
  on maps.  Altogether, there is a category $P\alg$ of $P$-algebras
  and $P$-algebra homomorphisms.
    
  {\em Lambek's lemma} says that if the category of $P$-algebras has an initial
  object $(A,a)$, then the structure map $a$ is invertible.  (This is not a 
  difficult result.)  This states precisely that an initial $P$-algebra
  $(A,a)$ is a solution to the equation $X \isopilback P(X)$: the underlying set
  $A$ is $X$ and the structure map $a$ is the required bijection.  Initiality
  is the technical condition that justifies reference to this as the {\em least
  fixpoint}.
  
  Now, the equation we wish to solve is not $X \isopilback P(X)$, but rather 
  $$
  X\isopilback 1+P(X),
  $$
  so what we are looking for is not the initial
  $P$-algebra, but rather the initial $(1+P)$-algebra.    This is a subtle 
  point; let us simply remark that the $1$ appears in the Dyson--Schwinger
  equations \eqref{eq:BK-DSE},
  and also that the $1$ in the polynomial fixpoint equations \eqref{eq:polyfix}
  has some strong motivations in category theory.  Its  presence ensures
  that the solution yields a nice class of trees, as discussed below.
\end{blanko}

\begin{theorem}
  If $P$ is a polynomial functor,
  then the fixpoint equation $$
  X\isopilback 1+P(X)
  $$ has a least solution, that is,  the category of $(1+P)$-algebras has
  an initial object.  This solution is the set of (isomorphism classes of) 
  $P$-trees, now to be defined.
\end{theorem}

This result is mostly folklore.  The explicit characterisation of the solution 
is from \cite{Kock:0807}.

\begin{blanko}{Operadic trees.}\label{operadictrees}
  By {\em operadic trees} we mean rooted trees admitting open-ended edges
  (leaves and root), such as the following:
  $$
    \begin{texdraw}
  \linewd 0.8 \footnotesize
  \move (-50 0)
  \bsegment
    \move (0 0) \lvec (0 20)
  \esegment
  
  \move (0 0)
  \bsegment
    \move (0 0) \lvec (0 10) \onedot
  \esegment
  
  \move (50 0)
  \bsegment
    \move (0 0) \lvec (0 26)
    \move (0 13) \onedot
  \esegment
  
  \move (105 0)
  \bsegment
    \move (0 0) \lvec (0 10) \onedot
    \lvec (-5 25) \onedot \lvec (-10 40)
    \move (0 10) \lvec (-14 21) \onedot
    \move (0 10) \lvec (5 25) \onedot \lvec (0 40)
    \move (5 25) \lvec (10 40)
    \move (0 10) \lvec (23 36)
  \esegment
  \end{texdraw}
  $$
  They are called operadic because each node is regarded
  as an operation, with the incoming edges (reading from top to bottom)
  as input slots and the outgoing edge as output slot.  Note the
  difference between a leaf (an open-ended edge) and a nullary node.
\end{blanko}

\begin{blanko}{$P$-trees.}
  (Cf.~\cite{Kock:0807}) For $P$ a polynomial functor represented by a set map
  $p:E\to B$, we think of the elements in $B$ as operations.  The most important
  aspect of $b$ is its {\em arity}, which is not just a number, but rather the set
  $E_b$ itself, interpreted as the {\em set} of input slots of the
  operation $b$.  For example, if $E_b$ is a $2$-element set,  $b$ is a
  binary operation.  It is sensible to picture the elements in $B$ as
  corollas:
  
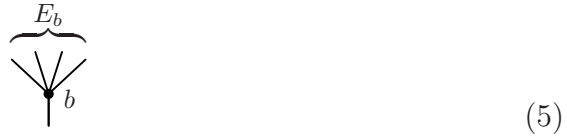
\begin{equation}\label{tree-draw}
    \begin{texdraw}
  \linewd 0.8 \footnotesize
  \move (0 0)
  \bsegment
    \move (0 0) \lvec (0 12) \onedot 
    \lvec (-14 25)
    \move (0 12) \lvec (-5 28) 
    \move (0 12) \lvec (5 28) 
    \move (0 12) \lvec (14 25)
    \move (8 10) \htext {$b$}
    \move (0 42) \htext {$E_b$}
    \move (0 32) \htext {$\overbrace{\phantom{xxxxx}}$}

  \esegment
  \end{texdraw}\end{equation}
  
  A {\em $P$-tree} is an operadic tree with nodes decorated by elements in $B$,
  and for each node $x$ decorated by $b$ a specified bijection between the
  incoming edges of $x$ and the set $E_b$.  In other words, each node is
  decorated with an operation of matching arity, and hence a $P$-tree can also
  be regarded as a tree configuration of operations from $P$.
\end{blanko}

\begin{blanko}{Example: Binary trees.}\label{ex:binary}
  Consider the polynomial functor $P$ defined by the set map $p:
  \{\mathtt{left},\mathtt{right}\} \to 1$.  It is the functor
  \begin{eqnarray*}
    \Set & \longrightarrow & \Set  \\
    X & \longmapsto & X^2 .
  \end{eqnarray*}
  For this $P$, a $P$-tree is precisely a (planar) binary tree.
  Indeed, since in this case the set $B$ is just singleton,
  to $P$-decorate a tree amounts to specifying a bijection, for each
  node, between the set of incoming edges and the set 
  $\{\mathtt{left},\mathtt{right}\}$.  For this bijection to be possible,
  each node must have precisely two incoming edges, and the bijection says
  which is the left branch and which the right.

  The relevant fixpoint equation $X\isopilback 1+P(X)$ is now
  $$
  X \isopilback 1 + X^2 ,
  $$
  and
  the theorem thus says that the solution, the initial $(1+P)$-algebra, is the set
  of planar binary trees.  Indeed, the fixpoint equation can be read as saying:
  a planar binary tree is either the trivial tree, or it is given by a pair of
  planar binary trees.  This is precisely the recursive characterisation of
  binary trees.
  Here are the first few elements:

  \begin{equation}\label{eq:binary-trees}\begin{texdraw}\setunitscale 0.8
    \move (0 0) 
    \bsegment
      \rlvec (0 20) 
    \esegment
    
    \move (40 0) 
    \bsegment
      \move (0 0)
      \lvec (0 10) \onedot \rlvec (-5 10)
      \move (0 10) \rlvec (5 10)
    \esegment
    
    \move (85 0)   
    \bsegment
      \move (0 0)
	\lvec (0 10) \onedot \rlvec (-10 20)
        \move (0 10) \rlvec (10 20)
        \move (-5 20) \onedot \rlvec (5 10)
    \esegment

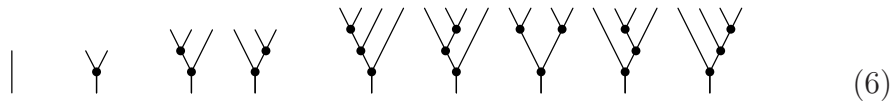
    \move (115 0)   
    \bsegment
      \move (0 0)
	\lvec (0 10) \onedot \rlvec (-10 20)
        \move (0 10) \rlvec (10 20)
        \move (5 20) \onedot \rlvec (-5 10)
    \esegment
    
    \move (170 0)
    \bsegment
      \move (0 0) 
      \bsegment
	\move (0 0) 
	\lvec (0 10) \onedot \rlvec (-15 30)
	\move (0 10) \rlvec (15 30)
	\move (-5 20) \onedot \rlvec (10 20)
	\move (-10 30) \onedot \rlvec (5 10)
      \esegment
     
      \move (40 0) 
      \bsegment
	\move (0 0) 
	\lvec (0 10) \onedot \rlvec (-15 30)
	\move (0 10) \rlvec (15 30)
	\move (-5 20) \onedot \rlvec (10 20)
	\move (0 30) \onedot \rlvec (-5 10)
      \esegment
   
      \move (80 0) 
      \bsegment
	\move (0 0) 
	\lvec (0 10) \onedot \rlvec (-15 30)
	\move (0 10) \rlvec (15 30)
	\move (-10 30) \onedot \rlvec (5 10)
	\move (10 30) \onedot \rlvec (-5 10)
      \esegment
     
      \move (120 0) 
      \bsegment
	\move (0 0) 
	\lvec (0 10) \onedot \rlvec (-15 30)
	\move (0 10) \rlvec (15 30)
	\move (5 20) \onedot \rlvec (-10 20)
	\move (0 30) \onedot \rlvec (5 10)
      \esegment

      \move (160 0) 
      \bsegment
	\move (0 0) 
	\lvec (0 10) \onedot \rlvec (-15 30)
	\move (0 10) \rlvec (15 30)
	\move (5 20) \onedot \rlvec (-10 20)
	\move (10 30) \onedot \rlvec (-5 10)
      \esegment
      
    \esegment
  \end{texdraw}\end{equation}
%
%
\end{blanko}

%
%

%

\begin{blanko}{Example: Planar trees.}\label{ex:planar}
  Take $P(X) = X^0 + X^1 + X^2 + X^3 + \cdots$ This is the list
  endofunctor, which sends a set $X$ to the set of lists of elements in $X$.
  Then $P$-trees are planar trees.
  The fixpoint equation 
  $$
  X \isopilback 1 + X^0 + X^1 + X^2 + X^3 + \cdots
  $$
  says that a planar tree is either the trivial tree or a list of planar trees.

  Since we allow nullary and unary nodes (corresponding to the terms $X^0$ and
  $X^1$ in the polynomial functor), for each fixed number of leaves, there are
  infinitely many trees.  For the sake of comparison with \ref{BK-DSE}, it is
  interesting to tweak this functor slightly in order to avoid this infinity:
\end{blanko}
  
\begin{blanko}{Example: Stable planar trees.}\label{ex:stable}
  Consider instead the polynomial functor 
  $$P(X) = X^2 + X^3 +X^4 + \cdots$$
  for which $P$-trees are {\em stable} planar trees, meaning they have no nullary or 
  unary nodes.  The exclusion of nullary and 
  unary nodes implies that, for each fixed number $k$, there is now only
  a finite number of trees with $k$ leaves. 
  These are the 
  Hipparchus--Schr\"oder numbers, $1, 1, 3, 11, 45, 197, 903,\ldots$
  Here are pictures of all the stable
  planar trees with up to $4$ leaves:
   
    \begin{equation}\label{eq:stable-trees}\begin{texdraw}
      \setunitscale 0.8
    \move (0 0) 
    \bsegment
      \rlvec (0 20) 
    \esegment
    
    \move (80 0) 
    \bsegment
      \move (0 0)
      \lvec (0 10) \onedot \rlvec (-5 10)
      \move (0 10) \rlvec (5 10)
    \esegment
    
    \move (160 -5)   
    \bsegment
      \move (0 0)
          \bsegment
	    \move (0 0)
	    \lvec (0 10) \onedot \rlvec (-10 20)
	    \move (0 10) \rlvec (10 20)
	    \move (-5 20) \onedot \rlvec (5 10)
	  \esegment

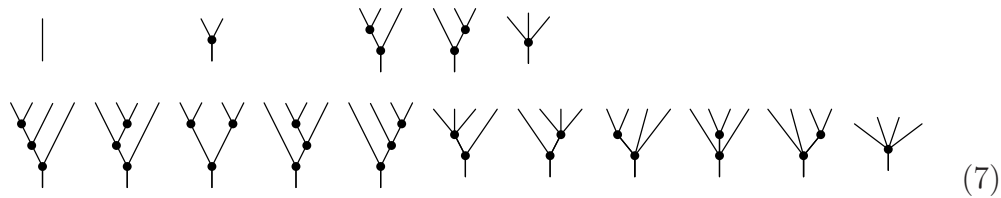
	  \move (35 0)
	  \bsegment
	    \move (0 0)
	    \lvec (0 10) \onedot \rlvec (-10 20)
	    \move (0 10) \rlvec (10 20)
	    \move (5 20) \onedot \rlvec (-5 10)
	  \esegment
	  \move (70 0)
	  \bsegment
	    \move (0 4)
	    \lvec (0 14) \onedot \rlvec (-10 12)
	    \move (0 14) \rlvec (10 12)
	    \move (0 14) \rlvec (0 14)
	  \esegment
    \esegment
    
    \move (0 -60)
    \bsegment
      \move (0 0) 
      \bsegment
	\move (0 0) 
	\lvec (0 10) \onedot \rlvec (-15 30)
	\move (0 10) \rlvec (15 30)
	\move (-5 20) \onedot \rlvec (10 20)
	\move (-10 30) \onedot \rlvec (5 10)
      \esegment
     
      \move (40 0) 
      \bsegment
	\move (0 0) 
	\lvec (0 10) \onedot \rlvec (-15 30)
	\move (0 10) \rlvec (15 30)
	\move (-5 20) \onedot \rlvec (10 20)
	\move (0 30) \onedot \rlvec (-5 10)
      \esegment
   
      \move (80 0) 
      \bsegment
	\move (0 0) 
	\lvec (0 10) \onedot \rlvec (-15 30)
	\move (0 10) \rlvec (15 30)
	\move (-10 30) \onedot \rlvec (5 10)
	\move (10 30) \onedot \rlvec (-5 10)
      \esegment
     
      \move (120 0) 
      \bsegment
	\move (0 0) 
	\lvec (0 10) \onedot \rlvec (-15 30)
	\move (0 10) \rlvec (15 30)
	\move (5 20) \onedot \rlvec (-10 20)
	\move (0 30) \onedot \rlvec (5 10)
      \esegment
       
      \move (160 0) 
      \bsegment
	\move (0 0) 
	\lvec (0 10) \onedot \rlvec (-15 30)
	\move (0 10) \rlvec (15 30)
	\move (5 20) \onedot \rlvec (-10 20)
	\move (10 30) \onedot \rlvec (-5 10)
      \esegment

      \move (200 0) 
      \bsegment
	\move (0 5) 
	\lvec (0 15) \onedot \rlvec (15 22)
	\move (0 15) \lvec (-5 25) 
	\onedot \rlvec (-10 12)
	\move (-5 25) \rlvec (10 12)
	\move (-5 25) \rlvec (0 12)
      \esegment

      \move (240 0) 
      \bsegment
	\move (0 5) 
	\lvec (0 15) \onedot \rlvec (-15 22)
	\move (0 15) \lvec (5 25) 
	\onedot \rlvec (-10 12)
	\move (5 25) \rlvec (10 12)
	\move (5 25) \rlvec (0 12)
      \esegment

      \move (280 0) 
      \bsegment
	\move (0 5) 
	\lvec (0 15) \onedot \lvec (-8 25) \onedot
	\rlvec (-5 12) \move (-8 25)
	\rlvec (5 12)
	\move (0 15) \lvec (6 37) 
	\move (0 15) \lvec (17 37) 
      \esegment
      
      \move (320 0) 
      \bsegment
	\move (0 5) 
	\lvec (0 15) \onedot \lvec (-14 37)
	\move (0 15) \lvec (14 37)
	\move (0 15) \lvec (0 25) \onedot
	\rlvec (-5 12) \move (0 25)
	\rlvec (5 12)
      \esegment
      
      \move (360 0) 
      \bsegment
	\move (0 5) 
	\lvec (0 15) \onedot \lvec (8 25) \onedot
	\rlvec (5 12) \move (8 25)
	\rlvec (-5 12)
	\move (0 15) \lvec (-6 37) 
	\move (0 15) \lvec (-17 37) 
      \esegment

      \move (400 0) 
      \bsegment
	\move (0 8) 
	\lvec (0 18) \onedot 
	\lvec (-16 30)
	\move ( 0 18) \lvec (-5 33)
	\move ( 0 18) \lvec (5 33)
	\move ( 0 18) \lvec (16 30)
      \esegment

    \esegment
    
    \setunitscale 1
  \end{texdraw}\end{equation}
%
%
\end{blanko}

\bigskip

We briefly list some further facts 
to illustrate the workings of $P$-trees, and highlight a few of
the features of this approach to Dyson--Schwinger equations.
See \cite{Kock:Poly-DSE} for all details.

\begin{blanko}{Bialgebra of $P$-trees.}
   $P$-trees form a Connes--Kreimer-style bialgebra~\cite{Kock:1109.5785}.  Note
   however, that cut edges are really cut rather than removed, as
   exemplified by
  
  \vspace{-4pt}
  
  \begin{center}\begin{texdraw}
    \htext (0 0) {$\Delta($}
   \rmove (12 0) \twotree
     \rmove (15 0)
    \htext  {$) \ \ = $}
        \rmove (23 0) \zerotree
    \rmove (3.5 0) \zerotree
    \rmove (3.5 0) \zerotree

          \rmove (11 0) \htext {$\tensor$}
   \rmove (14 0) \twotree

	            \rmove (18 0) \htext {$+$}

    \rmove (20 0) \onetree  \rmove (5 0) \zerotree 
    
    \rmove (12 0) \htext {$\tensor$}
   \rmove (12 0) \onetree
   
      \rmove (19 0) \htext {$+$}

    \rmove (21 0) \twotree
    
    \rmove (13 0) \htext {$\tensor$}
   \rmove (10 0) \zerotree
  \end{texdraw}\end{center}
  This is an essential point, as otherwise the decorations would be spoiled:
  removing an edge rather than cutting it would break the given arity bijections
  in the decorations (not rendered in the drawing).  Note also that this bialgebra is not connected: the
  degree-zero piece is spanned by the nodeless trees and forests.
%

  Each $b\in B$ defines a $B_+$-operator (although {\em not} a Hochschild
  $1$-cocycle), in terms of which the original equation~\eqref{eq:polyfix} can be internalised
  to this bialgebra.
\end{blanko}

\begin{blanko}{Core.}\label{core}
  There is a canonical bialgebra homomorphism from any such bialgebra of
  $P$-trees to the Connes--Kreimer bialgebra $\CK$, given by taking
  core~\cite{Kock:1109.5785}: this amounts to forgetting the $P$-decorations and
  shaving off all leaf edges and the root edge.  In other words, the core of a
  $P$-tree is the combinatorial tree given by its inner edges.
  
  Consider the binary trees constituting the least fixpoint for $X \mapsto
  1+X^2$, the first few of which are listed in \eqref{eq:binary-trees}.  Taking
  core transforms these binary trees into sub-binary combinatorial trees (i.e., they have at most two incoming edges at each node), and where the planar
  structure has been lost.  We see that the coefficients $c_k$ appearing in the
  solution of the quadratic Dyson--Schwinger equation of \ref{ex:binary-DSE} are
  precisely the numbers of binary trees with a given core.  (This interpretation
  of the coefficients $c_k$ has been given already by Bergbauer and
  Kreimer~\cite{Bergbauer-Kreimer:0506190}, in some form.)

  Similarly, consider the stable planar trees from \ref{ex:stable}, the first few
  of which are listed in \eqref{eq:stable-trees}.  Taking core yields exactly the 
  combinatorial trees found in the solution to the Dyson--Schwinger equation
  in \ref{ex:infinite-DSE}, and again the coefficients $c_k$ in the solution
  are precisely the numbers of trees with $k+1$ leaves and a given core.
\end{blanko}

\begin{blanko}{Fa\`a di Bruno sub-bialgebras.}
  While taking core immediately places us in the realm of the familiar Connes--Kreimer Hopf
  algebra, it is  important that the information
  contained in the leaves and root is not discarded: the
  strict type obedience characteristic for $P$-trees 
  (respect for arities) allows for meaningful automorphism
  groups and the existence of meaningful Green functions
  $$
  G = \sum_{T} \frac{T}{\norm{\Aut(T)}},
  $$
  where the sum is over iso-classes of $P$-trees.
  (To actually observe any automorphisms, one must work with groupoids instead of 
  sets, cf.~\cite{Kock:MFPS28} and \cite{GalvezCarrillo-Kock-Tonks:1207.6404}, 
  but that is beyond the scope of the present exposition.)
  This sum can be split into summands given by trees with $n$ leaves,
  $$
  G = \sum_{n\geq 0} g_n.
  $$
  Further, there is now a Fa\`a di Bruno formula~\cite{GalvezCarrillo-Kock-Tonks:1207.6404}
  $$
  \Delta(G) = \sum_{n\geq 0} G^n \tensor g_n,
  $$
  in the style of van Suijlekom~\cite{vanSuijlekom:0807}.  The point here is that 
  the exponent $n$ on the left-hand tensor factor 
  counts $n$ trees, each with a root, precisely matching the 
  subscript $n$ in the right-hand tensor factor, which is the number of leaves on the 
  trees in $g_n$.  This kind of information cannot be observed at the level of 
  combinatorial trees.
 
  Concern may arise that $P$-trees form bialgebras and not Hopf algebras,
  as used in renormalization.  However, this is not a problem, as it can be
  shown~\cite{Kock:1411.3098} that Hopf-algebra renormalisation also works for
  bialgebras of $P$-trees.  In fact it is insinuated in \cite{Kock:1411.3098}
  that the bialgebras are closer to actual physics than the Hopf algebras that 
  can be derived from them.
\end{blanko}

\section{Type theory}

We now proceed to interpret the polynomial fixpoint equations in type theory,
our first task being to explain what type theory is about.  Doing this in just a
few pages will necessarily be a superficial account.  Specifically, we mostly neglect
the important notion of {\em identity types}, which is playing an increasingly
central role in modern research~\cite{HoTT-book} (see \ref{HoTT}).  In
particular, we use the equality symbol $=$ in the most naive manner.

A classic reference on this subject is
Nordstr\"om--Petersson--Smith~\cite{Nordstrom-Petersson-Smith:programming}.  A
more modern account, which is highly recommended, is the book {\em Homotopy Type
Theory---Univalent Foundations of Mathematics}~\cite{HoTT-book}.


Type theory provides a foundation for mathematics.  Before explaining its primary concepts,
we first take a very brief look at the most common foundation for 
mathematics, set theory.

\begin{blanko}{Set theory.}\label{sets}
  The standard foundation for mathematics is set theory, and more specifically 
  what is called Zermelo--Fraenkel set theory.  This theory begins with first-order logic,
  the language written with operators
  $$
  \begin{array}{|c|c|}
    \hline
    \wedge & \text{conjunction (\texttt{AND})}  \\
    \hline
    \vee & \text{disjunction (\texttt{OR})}  \\
    \hline
    \top & \texttt{TRUE}  \\
    \hline
    \bot & \texttt{FALSE}  \\
    \hline
    \Rightarrow & \text{implication}  \\
    \hline
    \forall & \text{universal quantifier}  \\
    \hline
    \exists & \text{existential quantifier}  \\
    \hline
     
  \end{array}
  $$
  On top of this, Zermelo--Fraenkel set theory is defined as a one-sorted theory with one binary 
  relation, namely the membership relation $\in$, as in $a\in A$, expressing that
  $a$ is a member of $A$.  Here, both $a$ and $A$ are sets --- the
  only kind of object there is (that's what `one-sorted' means).
  
  Depending on how it is formulated, there are about ten axioms, all written
  in first-order logic.  Most of these axioms express what is needed for the
  elementary theory of sets, as used in everyday
  mathematics, and say for example that there exists the `empty set'
  $\emptyset=\{\}$, that one can add a new element to a given set, that one
  can form the union of two sets, that the 
  subsets of a given set form a set again, that there exist infinite sets, and
  so on.  In addition, there are some further, more technical axioms, 
  which are primarily useful for
  the application of set theory to encode all of mathematics.   For many purposes, the axiom of 
  choice is also added in some form.
  
  Zermelo--Fraenkel set theory serves as a foundation for mathematics in the sense that all of
  mathematics can be encoded in it, beginning with the natural numbers
  and progressing to the rationals, the reals, all of algebra, analysis, geometry, and so on.
  However, although this encoding can be achieved, there is no really canonical way of doing it.
  For example, here are two different ways of encoding the natural numbers.
  Von Neumann did it in this way: define $0 = \{\}$, the empty set, and define 
  successively each new natural number to be the set of all the previously defined 
  natural numbers: 
  $$1= \{0\} =\{\{\}\}, 2= \{0,1\}=\{ \{\}, \{\{\}\} \},
  3=\{0,1,2\}= \{ \{\}, \{\{\}\} , \{ \{\}, \{\{\}\} \} \},\ldots
  $$
  
  Zermelo did like this:  
  define $0 = \{\}$, the empty set, and successively define 
  inductively each new natural number to be the set containing only the previous
  natural number:
  $$
  0= \{\}, 1=\{0\}= \{\{\}\} , 2 = \{1\} = \{\{\{\}\}\}, 3= \{2\} = 
  \{\{\{\{\}\}\}\}, \ldots
  $$
  
  Everything in mathematics {\em can} be encoded as sets, but since this
  encoding is somewhat arbitrary or artificial, it is also possible to write things
  that actually make no sense mathematically.  For example, $3\in 17$.  This is a
  proposition in Zermelo-Fraenkel set theory, which is to say that it is either
  true or false.  However, its truth value depends on how the natural numbers
  are encoded. For example, this proposition is true according to von Neumann's
  encoding and false according to the Zermelo encoding.  (This is a famous
  example due to Benacerraf (1965).)  This is regarded by some as a conceptual
  problem with Zermelo-Fraenkel set theory: although it can encode all of
  mathematics, its structure does not faithfully reflect the structure of
  mathematics.
\end{blanko}

\begin{blanko}{Type theory.}
  Type theory is an alternative to set theory as a foundation for mathematics.
  Its development was initiated by Russell in the 1930s, and found its modern form
  mainly in the work of Martin-L\"of in the 1970s and 1980s.  What we loosely
  refer to as type theory is more precisely called {\em dependent type theory}
  or {\em Martin-L\"of type theory}, but we completely gloss over the finer 
  details that distinguish its many variants.  
  
  The basic ingredients in type theory are type declarations (called judgements),
  such as 
  $$
  a:A,
  $$ 
  akin to programming languages where one might write \texttt{n:INT} to mean
  that $n$ is a variable of integer type.  Here it reads $a$ is a term of type
  $A$.  Before writing it, one should actually declare that $A$ is a type:
  $$
  A:\Type
  $$
  (This can be regarded as a special kind of the basic type declaration,
  provided one assumes a universe type $\Type$ of all types.)
  (The word `declaration' has the advantage that it reminds us that $a:A$
  is not a proposition: for example, it cannot be negated.  In this way, it differs
  from the operator $\in$ in Zermelo--Fraenkel set theory, where $a\in A$ is a 
  proposition, i.e., it may or may not be true.) 

  
  Type theory was developed to provide a {\em constructive} foundation for
  mathematics, and as a result, it has proved very useful as a foundation for
  computer science.  In fact, type theory can be regarded as a programming
  language, and everything written in type theory can be verified using proof
  assistants such as Coq~\cite{Coq} or Agda~\cite{Agda}.  
  With the recent advent of Homotopy Type Theory and Univalent
  Foundations~\cite{HoTT-book} (whose main points are outside the scope of this
  exposition), the aim of providing a practical foundation for mathematics is
  coming nearer fulfilment, with a closer interaction between humans and
  computers as an important bonus feature.
\end{blanko}

\begin{blanko}{Sets as types, and some basic type formers.}
  At a first level of understanding, a judgement $a:A$ can be read as a
  membership, $a\in A$.  There are constructions and rules in type theory
  allowing one to do elementary set theory in this way.

  For example, one can form product types: given that $A:\Type$ and $B:\Type$
  then there is inferred a new type $A\times B:\Type$.  The inference rules in type
  formation like this are traditionally written with a horizontal bar:
  $$
  \frac{A:\Type,\  B:\Type}{A\times B:\Type}
  $$
  meaning that if the judgements above the line are assumed, then the judgements
  below it can be inferred.  (A more informal writing idiom, more akin to prose, 
  is advocated in
  the book \cite{HoTT-book}.)
  There is also a singleton type $1:\Type$.

  Similarly, there is a function type, denoted $A\to B$:
    $$
  \frac{A:\Type,\  B:\Type}{A\to B: \Type}
  $$
  A term of this type, $f: A \to B$, is the analogue of an actual function
  or set-mapping in set theory.
  There are further rules expressing how to construct terms of these new types,
  and how to compute with them.
\end{blanko}
  
\begin{blanko}{Dependent types.}
  An important aspect is {\em dependent types}.  These are families of types
  indexed by a given type.  With the universe interpretation of $\Type$, a 
  dependent type can
  be written as
  $$
  B:A \to\Type 
  $$
  It says that for each term $a:A$, there is a type $B(a)$.
  The analogue in set theory is a family of sets $\{B(a)\mid a\in A\}$
  indexed by a set $A$, as in \eqref{eq:setfamily}. 
  For what follows, it is important to see this also 
  as being encoded as a single map of sets $B \to A$, often thought of as a fibration:
  then each family member
  $B(a)$ is defined to be
  the fibre over a point 
  $a\in A$.  Conversely, given a family of sets like this, one can form
  the map $\sum_{a\in A} B(a) \to A$, where the sum sign signifies a disjoint 
  union of sets, and the map is projection onto the indexing set of the sum.
\end{blanko}

\begin{blanko}{Dependent products.}
  We have already mentioned function types: a term of type $A\to B$
  is interpreted as an assignment that to every term in $A$ associates a term
  in $B$.  Here $A$ and $B$ are fixed types.  There is a dependent version,
  which will be crucial in what follows.
  Suppose $B: A \to\Type$ is a dependent type.
  Then there is a new type $\prod_{a:A} B(a)$ called {\em dependent product}:
  $$
  \frac{B: A \to\Type}{\prod_{a:A} B(a)}
  $$
  A term of this type is again interpreted as an assignment that to every term $a:A$
  associates a term, but this time in a `codomain' that depends on $a$ itself.
  Under the interpretation of the dependent type $B:A \to\Type$ as a fibration
  $p:B \to A$, the terms in $\prod_{a:A}B(a)$ are sections to $p$.
\end{blanko}


\begin{blanko}{Propositions as types.}
  We have seen that type theory can emulate elementary set theory by interpreting
  types as sets and terms as elements; then type declaration is read as membership.
  However, there is another very useful interpretation of type theory, which
  shows that type theory {\em contains} all of first-order logic, rather than 
  depending on it as a meta language.  In this
  interpretation, a type is interpreted as a proposition and a term is
  interpreted as a proof.  Hence the type declaration $a:A$ says that $a$ is a
  proof of $A$.  This ties in with the constructive aspect of type theory: to
  establish the truth of a proposition, one must explicitly {\em construct} a
  proof term using the rules of type theory.

The basic type formers (not all of which were explained here)
then have the following logical interpretation:

  $$
  \begin{array}{|c|c|}
    \hline
    \times &\wedge  \\
    \hline
    + & \vee  \\
    \hline
    1& \top   \\
    \hline
    0 & \bot   \\
    \hline
    \to &\Rightarrow   \\
    \hline
    \prod & \forall   \\
    \hline
    \sum & \exists   \\
    \hline
     
  \end{array}
  $$
It should be noted that the strict interpretation of $1$ as `true' actually 
means rather
`true with a unique proof.'  More loosely, when a type is interpreted as a 
proposition, any term of that type will be interpreted as a proof of the 
corresponding proposition, so a looser notion of true is `inhabited,' or as 
most people would put it, `non-empty.'  This is the notion used in what follows.
\end{blanko}

\begin{blanko}{Predicates as dependent types.}
  In logic, a {\em predicate} is a proposition that depends on a variable.
  Under the propositions-as-types interpretation, the corresponding notion is
  precisely dependent types.  This interpretation can be very helpful for the reading of
  complicated formulae, particularly when it comes to dependent products, which
  are then read as universal quantifiers.  Recall that we had previously
  viewed the dependent product $\prod_{a:A} B(a)$ as a kind of function space, a
  space of sections: for {\em every} $a$ in $A$ we assign some $f(a)$ in $B(a)$.
  If now instead we interpret $B(a)$ as a predicate, then a term in
  $\prod_{a:A}B(a)$ is a proof that {\em for all} $a$ in $A$ the predicate
  $B(a)$ holds.  Similarly, although we shall not really need it here, the notion of dependent
  sum corresponds to the existential quantifier.
\end{blanko}

\begin{blanko}{Type formation rules.}
  We have, very superficially, mentioned many type formers: cartesian product,
  singleton type, function types, dependent products, and dependent sums.  Each
  time, we sketched what they are, but neglected to list the further rules
  governing them.  In fact, one beauty of type theory is that these rules follow
  a very strict pattern, common for all type formers.  We explain this
  pattern here, and then exemplify it in the following sections, where we exploit it in
  more detail to describe inductive types.
  
  Every type former is given by four rules (or four groups of rules).
  
  The first is a {\em formation rule}, which stipulates the new type.
  For the product type, this just reads
  $$
  \frac{A:\Type, \ B:\Type}{A\times B:\Type}
  $$
  (So far in the text, we have only seen formation rules.)
  
  The second rule, or group of rules, is called the {\em introduction rule}:
  it populates the new type with terms that characterises it.  In the case 
  of the product, it says that the new terms are pairs of terms:
  $$
  \frac{a:A, \ b:B}{\mathsf{pair}(a,b):A\times B}
  $$
  The individual introduction rules are referred to as {\em constructors}.
  The product type is thus characterised by a single constructor, 
  $\mathsf{pair}$.
  
  Next comes the {\em elimination rule}, which tells how terms in the type
  are used; that is, lists their characteristic properties by stipulating an
  {\em eliminator}, as we shall see in the examples below.
  Finally there is the {\em computation rule} which stipulates how introduction 
  and elimination interact.
  The elimination and computation rules tend to appear a bit unwieldy, but we
  shall see in detail in the next sections how they work.  
\end{blanko}

\section{Classical induction: The natural numbers}
\label{sec:N}

Since type theory is constructive, practically the only way of getting
hold of infinite structures is through induction principles.
They include first of all the natural numbers, the most basic 
inductive type, but also trees of many kinds, as we shall see
in Section~\ref{sec:W}.

%
%

\begin{blanko}{Dedekind--Peano natural numbers.}
  Dedekind (1888) and Peano (1889) defined (or rather characterised)
  the natural numbers as a set $\N$ with
  a distinguished element $0\in \N$ and a successor function $s: \N
  \to \N$ satisfying

  \begin{punkt-i}
    \item
  $0$ is not a successor;

  \item
  Every element $x\neq 0$ is a successor;

  \item
  The successor function is injective;

  \item
  If a subset $U \subset \N$ contains $0$ and is stable under the
  successor function, then $U = \N$.
\end{punkt-i}

  Note that (i)+(ii)+(iii) amount to saying that the map
  $$
  \{*\} + \N \stackrel{\langle 0,s \rangle}{\longrightarrow} \N
  $$
  is a bijection.
  Axiom (iv) is called the {\em induction axiom}.
  
  Note that unlike Zermelo's and von Neumann's definitions briefly mentioned in
  \ref{sets} above, the Dedekind--Peano definition does not actually define a
  concrete set in terms of its elements.  Rather, the set is defined {\em
  structurally}, meaning that it is characterised by how it works, through its
  relationship with other sets.  In fact, each of
  Zermelo's
  and von Neumann's definitions can be seen as a specific implementation of 
  the Dedekind--Peano axioms.
\end{blanko}

\begin{blanko}{Lawvere: Natural numbers as an initial algebra.}
  Lawvere (1964) observed that the Dedekind--Peano definition can be
  reformulated categorically as an example of an initial algebra, as follows.
  The set of natural numbers is a set $\N$ together with an element $0\in\N$ and
  a map $s: \N \to \N$, with the following property: whenever $A$ is a set with
  $a \in A$ and $r: A \to A$, there is a unique function $u: \N \to A$ such
  that $u(0) = a$ and $u(s(n)) = r(u(n)), \forall n \in\N$.  Phrased in
  the terminology of Section~\ref{sec:poly}, the data given (the zero and the
  successor function) amount precisely to saying that $\N$ is an algebra for the
  polynomial functor $X \mapsto 1+X$.  And the characterising property says
  precisely that for any other such algebra $A$, there is a unique homomorphism
  of algebras $\N \to A$.  Put differently, the Peano--Dedekind axioms for the
  natural numbers say precisely that $\N$ is the initial algebra for the
  polynomial functor $X \mapsto 1+X$!  (In other words, the natural numbers are
  $P$-trees, for $P$ the identity functor.)
  
  (Lawvere's observation was made in the context of his {\em Elementary
  Theory of the Category of Sets} (see \cite{Lawvere-Rosebrugh}), which is a
  {\em structural} alternative to Zermelo--Fraenkel set theory: instead of
  starting from the membership relation, it takes the notion of maps of sets as
  primitive.  A set is then no longer characterised by its elements, but rather by its
  relationship to other sets.  While this may seem abstract at first sight, it
  is actually much closer than Zermelo--Fraenkel set theory to mathematical
  practice.  The Dedekind--Peano--Lawvere definition of the natural numbers
  gives a hint of the flavour of this structural set theory.)
\end{blanko}

\begin{blanko}{Natural numbers in type theory.}
  In type theory, the natural numbers are introduced as a type, as follows.
  
The formation rule simply stipulates that there is a type $\N$:
$$
\frac{}{\N:\Type}
$$

The introduction rule specifies the two constructors
$$
\frac{}{\mathsf{zero} : \N }
\quad \quad \frac{n:\N}{\mathsf{succ}(n): \N}
$$
(The second can be thought of as a $B_+$-operator, with reference to 
Example~\ref{ex:linear-DSE}: it takes a `forest' consisting 
of a single (unary) tree and returns a new (unary) tree by grafting that
tree onto a new root node.)


The elimination rule is where the induction principle is encoded.
The correct version involves {\em dependent elimination}.  Before coming to it,
it is worth giving a simplified version,  the non-dependent elimination rule,
which is easier to grasp, but which is not quite sufficient.
Trying to mimic what the initiality means, we are led to write
$$
\frac{A:\Type ,\quad  z:A,\quad  s:A\to A}{\mathsf{rec}:\N \to 
A}
$$
This is meant to say: given another such structure (i.e.~another type $A$ with the 
same constructors), there exists a map to it from $\N$.
And then finally write the computation rule, which states
that this function $\mathsf{rec}:\N \to A$ of course must be required to be compatible with 
the structure:
$$
\frac{\asbefore}{\mathsf{rec}(\mathsf{zero})=z , \quad
\mathsf{rec}(\mathsf{succ}(n))=s(\mathsf{rec}(n))  }
$$
(The symbol \asbefore, here and in what follows, is meant to say
`same hypotheses as in the elimination rule,' as this is always the case.)
The essence of all this is that $\N$ is designed so that we can define functions
out of it by recursion (hence the symbol $\mathsf{rec}$).

The missing bit in order to faithfully render the initial-algebra idea is
that we need to say that $\mathsf{rec}$ is {\em unique} with these properties.  This
is not something one can say directly in type theory, for reasons related
to its constructive nature.  What turns out to work much better is the
following {\em dependent elimination rule}, which we first write down, then explain
in detail:
\begin{equation}\label{eq:elimN}
  \frac{C:\N\to \Type ,\quad  z: C(0),\quad  
  s: \prod_{n:\N} C(n)\to C(\mathsf{succ}(n))}{\mathsf{rec}: \prod_{n:\N} C(n)}
\end{equation}

And finally the computation rule:
$$
\frac{\asbefore}{\mathsf{rec}(\mathsf{zero})=z\ ,\quad \mathsf{rec}(\mathsf{succ}(n)) = s(n,\mathsf{rec}(n))}
$$
\end{blanko}

\begin{blanko}{How to read (and write) elimination rules.}
  The non-dependent elimination rule says that under certain hypotheses,
  we can get a function out of our new type, in this case $\N$.
  The dependent elimination rule gives instead a dependent product.
  (Recall that a function type can be viewed as a `constant' dependent product.)
  A recommended way to read (and write) an elimination rule is to start with
  the following question:
$$
\frac{C:\N\to\Type ,\quad 
\begin{texdraw} \lpatt (1 3)
\move (0 0) \lvec (120 0) \lvec (120 14) \lvec (0 14) \lvec (0 0) \move (60 7)\htext{???} 
\end{texdraw}}
{\mathsf{rec}: \prod_{n:\N} C(n)}
$$
  The question asks: given a dependent type $C: \N \to \Type$, what is
  the data needed \raisebox{-2pt}{\begin{texdraw} \lpatt (1 3)
\move (0 0)\lvec (40 0)\lvec (40 12)\lvec (0 12)\lvec (0 0)\move (20 6)\htext{???} 
\end{texdraw}}
in order to obtain a term in $\prod_{n:\N} C(n)$?
  If we think of the dependent type as a `fibration,'
  the question is: what is needed to get a section?
    $$
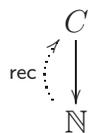
\xymatrix{
C \ar[d] \\
\N \ar@{..>}@/^1pc/[u]^{\mathsf{rec}}
}$$
  If instead we think of $C$ as a predicate, then the question is: what is needed
  to prove $C(n)$ for all $n$?

  Either way, once the question has been formulated, we proceed to fill the
  answer into the template to get the final rule \eqref{eq:elimN}, repeated here
  to stare at:
\begin{equation*}
  \frac{C:\N\to \Type ,\quad  z: C(0),\quad  
  s: \prod_{n:\N} C(n)\to C(\mathsf{succ}(n))}{\mathsf{rec}: \prod_{n:\N} C(n)}
\end{equation*}
The rule says: in
  order to construct a section $\mathsf{rec}$ we need: a point $z$ in the fibre
  over $0$, and a way of passing from one fibre to `the next': more precisely,
  for each $n$, we need a term of the function type $C(n) \to C(n+1)$.
  Alternatively, in terms of predicates: in order to prove $C(n)$ for all
  $n$, we must first prove $C(0)$ and then prove the `induction step':
  `assuming $C(n)$, prove $C(n+1)$.'
  If we have these ingredients, then we can deduce $C(n)$ for all $n$.

  Finally, there should be a computation rule, of course, telling us that 
  $\mathsf{rec}$ must be compatible with the original data given:
  $$
  \frac{\asbefore}{\mathsf{rec}(\mathsf{zero})=z\ ,\quad \mathsf{rec}(\mathsf{succ}(n)) = s(n,\mathsf{rec}(n))}
  $$
  (Notice that the function $s$ takes two arguments: the first is $n:\N$ (indexing
  the dependent product), the second is from $C(n)$, depending on the first 
  argument.)
  
  Note that the non-dependent eliminator is like defining functions
  recursively, while the dependent eliminator is like proof by induction.
  The former is a special case of the latter: given the abstract type $A$, one
  can always form the dependent type $n \mapsto A$ (the constant dependent
  type).  So given the dependent eliminator, we can emulate the
  non-dependent eliminator as a special case.  The dependent case is stronger,
  though: it actually implies the uniqueness.  For this we refer to the
  Appendix.
\end{blanko}

\begin{blanko}{Relationship with initial $(1+P)$-algebra (in this case, $P=\Id$).}
  Interpreting the dependent type as a `fibration,' the hypothesis of the
  elimination rule (the part above the bar) says precisely that there is a map
  $C \to \N$ and that this map is an $(1+P)$-algebra homomorphism.  The outcome
  of the elimination rule says that this map has a section, and finally the
  whole computation rule says that this section is in fact a $(1+P)$-algebra
  homomorphism.  So altogether, algebraically, the rules say that there is an
  algebra such that any algebra map into it has a section.  It is a general
  result in category theory (see Appendix) that this condition is equivalent to
  being an initial algebra.

  The link with combinatorial Dyson--Schwinger equations goes through the
  polynomial fixpoint equation.  There are two constructors: one nullary
  constructor stipulating that there exists a special element zero, and one
  unary constructor which takes as input one element and produces another (the
  successor).  The first corresponds to the $1$ in the Dyson--Schwinger equation,
  the second corresponds to the $B_+$-operator, and more precisely to the
  polynomial $X$ in its argument.
  The corresponding combinatorial Dyson--Schwinger equation is
  $$
  X = 1 + \alpha B_+(X) ,
  $$
  the ladder case \eqref{ex:linear-DSE}. 
\end{blanko}

\section{Inductive types: W-types }
\label{sec:W}

There is a quite general class of inductive types called {\em W-types}, W for
{\em wellfounded trees}, which as we shall see is closely related to
Dyson--Schwinger equations.  Before embarking on the general case, we go through
the example of binary trees.


\begin{blanko}{Binary trees.}
  We form the type of (planar) binary trees, here denoted $W$, following
  the same pattern as for the natural numbers, but with a binary constructor 
  instead of a unary constructor.

  Formation rule:
  $$\frac{}{W: \Type}
  $$
  Introduction rule:
  $$
  \frac{}{\mathsf{nil}:W}
  \quad\quad
  \frac{(t_1,t_2) : W\times W}{\mathsf{sup}(t_1,t_2):W}
  $$
  The second says that whenever we are given an ordered pair of trees,
  we can construct a new tree.
  It is traditionally called $\mathsf{sup}$ because it is in some manner the supremum of the two
  trees $t_1$ and $t_2$, namely the smallest tree containing $t_1$ and $t_2$
  as subtrees.  It corresponds precisely to the $B_+$-operator in 
  Example~\ref{ex:binary-DSE}.

  The crucial rule is the elimination rule.  (For the beginner, this is the most
  difficult rule to write down, but for inductive types there is actually a
  mechanical way of deriving it from the introduction rule.  For example, for 
  inductive types in the
  proof assistant Agda~\cite{Agda}, it is enough to write the formation and
  introduction rules, then the computer figures out by itself what the
  elimination and computation rules should be.)
%
%

  Elimination rule:
  $$
  \frac{C:W \to \Type , \ s_{\mathsf{nil}}: C(\mathsf{nil}), \ 
  s_{\mathsf{sup}}:
  \prod_{(t_1,t_2):W\times W} C(t_1)\times C(t_2) \to C(\mathsf{sup}(t_1,t_2))
  }{\mathsf{rec}:\prod_{w:W} C(w)}
  $$
%
%
%
  Let's go through it, in the interpretation of $C:W \to \Type$ as a predicate,
  i.e.~a proposition about (binary) trees.  The elimination rule then says: given a
  predicate on trees (that's the left-hand part of above-the-line) in order to
  prove this predicate for all trees (that's what's below the line), it is
  enough to be able to prove it for the trivial tree, and know how to derive
  from the statement about any two trees the statement about the tree obtained
  by grafting the two trees onto a new root node.

  And finally there is the computation rule:
  $$
  \frac{\asbefore}{\mathsf{rec}(\mathsf{nil}) = s_{\mathsf{nil}},\quad 
  \mathsf{rec}(\mathsf{sup}(t_1,t_2))  = 
  s_{\mathsf{sup}}(t_1,t_2,\mathsf{rec}(t_1),\mathsf{rec}(t_2))}
  $$
  
\end{blanko}

%
%
%
%
%

\begin{blanko}{W-types, general case.}
  See also \cite{HoTT-book}, \S5.3.  The notion of W-type is due to Martin-L\"of
  himself~\cite{MartinLof:1982}; it covers the two previous examples.
  The interpretation as initial algebras for
  polynomial functors is due to Moerdijk and
  Palmgren~\cite{Moerdijk-Palmgren:Wellfounded}.
  A W-type refers to a general dependent type
  $E : B \to \Type$,
  which plays the role of the polynomial (or power series)
  inside the $B_+$-operator in a combinatorial Dyson--Schwinger
  equation as in \ref{BK-DSE}.  Here $B$ is the set of possible branching types,
  and for fixed $b:B$, the type $E(b)$ is the arity of that node
  type.
  In the first example, that of natural numbers, we had
  $B=1$ (only one kind of node) and
  $E(1)=1$ (that node is unary).
  In the second example, that of binary trees,
  again $B=1$
  (only one kind of node) and now with
  $E(1)$ a $2$-element set (that node is binary).


  Relative to a given dependent type
  $E:B \to\Type$, we shall now define the W-type, to be thought of as
  the type of all trees of a certain kind.
  It ought to be denoted $W_{B,E}$ (or even $\mathrm{W}_{b:B} E(b)$) to
  express this dependency, but to lighten the notation we shall denote it just
  $W$.  Formation rule:
  $$
  \frac{E: B \to \Type}{W:\Type}
  $$
  Introduction rule:
  $$
  \frac{}{\mathsf{nil}:W} \quad \quad \frac{b:B , \quad t: E(b)\to 
  W}{\mathsf{sup}_b(t):W}
  $$
  In this case, it is appropriate to think of one $B_+$-operator for
  each element $b$ in $B$.  Each $b$ can be pictured as a small corolla
  $$
  \begin{texdraw}
  \linewd 0.8 \footnotesize
  \move (0 0)
  \bsegment
    \move (0 0) \lvec (0 12) \onedot 
    \lvec (-14 25)
    \move (0 12) \lvec (-5 28) 
    \move (0 12) \lvec (5 28) 
    \move (0 12) \lvec (14 25)
    \move (8 10) \htext {$b$}
    \move (0 42) \htext {$E(b)$}
    \move (0 32) \htext {$\overbrace{\phantom{xxxxx}}$}
  \esegment
  \end{texdraw}
  $$
  The rule says that given  an $E(b)$-indexed family of trees,
  we can glue all those trees onto the corresponding leaves of the corolla
  $b$ to obtain a new tree (with $b$ as its root node).

  The elimination rule now reads
  $$
  \frac{C:W\to \Type, \ s_{\mathsf{nil}}: C(\mathsf{nil}), \
  s_{\mathsf{sup}}: \red{\prod_{b:B}} \blue{\prod_{t:E(b)\to W}} \darkgreen{\prod_{e:E(b)} C(t(e))} \to 
  \magenta{C( \mathsf{sup}_b(t))}}{\mathsf{rec}: \prod_{w:W} C(w)}
  $$
  That's a mouthful, but the principle is exactly the same as for binary trees.
  It says that in order to prove something $C$ for this kind of trees,
  it is necessary to prove it for the trivial tree, and also establish
  the following induction step: 
  \red{for any kind of node $b$} (that's the `universal quantification' $\red{\prod_{b:B}}$),
  and \blue{for any $E(b)$-indexed family of trees}, assuming \darkgreen{$C$ holds for each of
  the trees in that family} (that's the $\darkgreen{\prod_{e:E(b)} C(t(e))}$ part),
  we can deduce \magenta{$C$ for the tree obtained by grafting onto $b$}.  Under these hypotheses
  we can then deduce that $C$ holds for all trees.

  And finally the computation rule:
  $$
  \frac{\asbefore}{\mathsf{rec}(\mathsf{nil})= s_{\mathsf{nil}}, \quad
  \mathsf{rec}(\mathsf{sup}_b(t)) =
  s_{\mathsf{sup}}(b,t,e,\mathsf{rec}(t(e)))}
  $$
\end{blanko}

\begin{rmk}
  In type theory, the nil constructor for the trivial tree is usually not listed
  separately, but is rather subsumed 
  as an extra nullary member of the dependent family.  
  For the present purposes it is preferable always to have
  this separate nil constructor, because on one hand it fits better into the
  polynomial formalism (where the initial $(1+P)$-algebra is the set of 
  operations of the free monad on 
  $P$), and also since the special term $1$ is included in the Dyson--Schwinger
  equations~\eqref{eq:BK-DSE}.  Foissy~\cite{Foissy:0707.1204} has studied 
  Dyson--Schwinger equations without this special term.  For a thorough analysis
  of the differences implied, see \cite{Kock:Poly-DSE}.
\end{rmk}

\section{Feynman graphs and outlook}

\begin{blanko}{Combinatorial trees versus graphs.}
  Ultimately, for the purposes of quantum field theory, 
  the combinatorial Dyson--Schwinger equations should take place in
  Hopf algebras of graphs, not of trees.  The interest in the Hopf
  algebra of combinatorial trees stems from the fact that it enjoys
  a universal property, which allows transfer of knowledge to the
  graph case.  Precisely, $\CK$ together with its
  canonical Hochschild $1$-cocycle $B_+$ can be shown to be an initial object in 
  the category of commutative combinatorial Hopf algebras equipped with a Hochschild 
  $1$-cocycle \cite{Connes-Kreimer:9808042}.
  In other words, for any other such Hopf algebra with a 
  $1$-cocycle, for example a Hopf algebra of Feynman graphs $\HH$, there is a unique
  Hopf algebra homomorphism $\CK\to\HH$ compatible with the Hochschild 
  $1$-cocycles.  In this way, $\CK$ serves as a universal approximation.
\end{blanko}

\begin{blanko}{$P$-trees versus graphs.}  
  $P$-trees provide more faithful approximations.  Given a class of
  graphs (belonging to some quantum field theory), it is possible to encode each
  graph as a tree decorated with primitive graphs, according to how the graph is
  built from nesting of primitive graphs.  It is shown in
  \cite{Kock:graphs-and-trees} that these trees can be described formally as
  $P$-trees for a suitable polynomial functor $P$, in such a way that the automorphism group of the
  $P$-tree agrees with the automorphism group of the graph with its nestings.
  The polynomial functors $P$ here are considerably more complicated than those
  considered in the present paper.  For one thing, since graph insertions
  must match residues with fertilities, the corresponding trees must have
  decorations also on the edges to keep track of such typing constraints.  This
  corresponds precisely to considering polynomial functors in many variables as
  in \cite{Kock:0807}.  Secondly, because of the existence of symmetries of graphs,
  it is necessary to upgrade the theory from polynomial functors over sets to
  polynomial functors over groupoids, as in \cite{Kock:MFPS28}.  (A groupoid is
  a category all of whose morphisms are invertible.)  Glossing over many
  details, the polynomial functor corresponding to a given quantum field theory
  is represented by the map $p:E \to B$, where $B$ is the groupoid of all
  primitive graphs of the theory (and their isomorphisms), and $E$ is the
  groupoid of all such graphs with a marked vertex (and their isomorphisms).
  There is now a bialgebra homomorphism $\HH\to\BB_P$ from the bialgebra of
  graphs to the bialgebra of $P$-trees.  It sends a graph to the sum of all the
  $P$-trees that are recipes for building it (that's a finite sum).  In other
  words, this bialgebra homomorphism precisely resolves overlapping divergences,
  and it does so in the gentlest possible way.  This bialgebra homomorphism
  is compatible with Green functions.  The $P$-trees form
  initial algebras --- they are genuine W-types, with all the good properties
  that entails.  The graphs themselves are {\em not} W-types, due precisely to
  the fact that some graphs can be constructed from primitive graphs in more
  than one way.
  
  It should be mentioned that some issues remain with regard to the
  polynomial-functor approach, which have not been sorted out satisfactorily yet
  (cf.~\cite{Kock:Poly-DSE} for further discussion).  The main issue is with
  edge insertions: the combinatorial Dyson--Schwinger equations in Hopf algebras
  of graphs typically involve denominators corresponding to propagators, and each
  edge represents an ordered infinity of insertion places. This feature is difficult
  to render in a strictly operadic setting, otherwise than insisting that all
  mass and kinetic terms be marked explicitly with crosses on the graphs.  This
  problem is under active investigation.
\end{blanko}
%
%


\begin{blanko}{Homotopy type theory: identity types as path spaces.}
  \label{HoTT}
  The above review of type theory is very crude.  One glaring omission is that
  of identity types.  In type theory, because of its strictly constructive nature,
  one cannot say that two terms of a type are equal without actually providing a proof,
  a {\em construction} of the equality, so to speak.
  This is handled
  as follows.  Given two terms $a$ and $b$ of a type $A$ there is a new
  type $\Id_A(a,b)$ called the {\em identity type}, whose terms can be thought 
  of as proofs
  that $a$ and $b$ are identical.  Formally there is a formation rule
  $$
  \frac{A:\Type, \quad a:A, \ b:A}{\Id_A(a,b): \Type}
  $$
  Now, two terms in the type $\Id_A(a,b)$ may or may not be equal, and if they are
  equal, that needs proof again, so there is a identity type of the identity type.  And
  so on.  {\em Homotopy type theory}~\cite{HoTT-book} exploits the significant
  discovery that this structure is intimately analogous to homotopies and higher
  homotopies in topology, so that a valid interpretation of types is to regard them as
  topological spaces (up to homotopy), terms are regarded as points, identity types are
  path spaces, and so on.
  In this manner, type theory can serve as a formal language for homotopy 
  theory.  
\end{blanko}
 
\begin{blanko}{Higher inductive types.}
  A fundamental ingredient in homotopy theory is the based circle $S^1$ (by
  which is meant any closed curve, from a point to itself).  For example,
  homotopy groups are defined by mapping based circles into spaces.  The circle
  can be rendered in homotopy type theory too~\cite{HoTT-book}, and is revealed
  to be of inductive nature as well. The circle is a basic example of a {\em higher
  inductive type}.  This means that it is specified in a way similar to W-types,
  but with constructors allowed to be terms not just in the type itself (called
  $0$-constructors) but also in its identity type (called $1$-constructors).
  Precisely, the circle $S^1$ is given by two constructors, namely a basepoint
  and a $1$-constructor which is a path from the basepoint to itself.  Formally,
  here is the introduction rule:
  $$
  \frac{}{\mathsf{base}:S^1} \quad \quad \frac{}{\mathsf{loop}: 
  \Id_{S^1}(\mathsf{base}, \mathsf{base})}
  $$
  
  The elimination and computation rules essentially follow the same pattern
  as we have seen for W-types, but they are slightly more involved.  The upshot is that
  the circle, as well as many other types of topological origin, which are
  also higher inductive types, can be manipulated `by induction.'
  
  It is worth mentioning here that even the notion of identity type itself is
  in some sense an inductive type, whose elimination and computation rules
  are analogous to those for W-types.  A remarkable range of structures
  are now susceptible to inductive methods.  For an outsider, the arguments
  employed may look like magic, but for the computer verifying them and to
  the seasoned programmer, they are simply a natural and absolutely fundamental
  principle taken to the next level of generality.
\end{blanko}

\begin{blanko}{Feynman graphs as higher inductive types?}
  While topologically a term in an identity type is a path, logically it is an
  equality, or an equation imposed.  While conventional W-types parametrise
  operations of free algebraic structures (technically, they are the operations 
  of the free monad on $P$~\cite{Kock:0807}), higher inductive types parametrise
  operations of free algebraic structures quotiented by certain equations.
    
  As explained above, graphs-with-a-nesting are precisely certain $P$-trees (and hence a
  conventional W-type), and graphs themselves are obtained by quotienting the
  set of $P$-trees by identifying two $P$-trees if they build the same graph.
  On these grounds, it is not unreasonable to speculate that  Feynman
  graphs may form a higher inductive type.
  Actually establishing this will require further structural insight into
  Feynman graphs.
 
  Higher inductive types are a rather recent addition to type theory, and their
  formalisation has not yet completely crystallised.  In particular, the
  appropriate freedom in imposing equations has not yet been fully determined.
  It transpires from work of Lumsdaine and
  Shulman~\cite{Lumsdaine-Shulman:semantics-HIT} that the specification of the
  equations ($1$-constructors) should be `polynomial' in nature, in a certain
  sense, just as the $0$-constructors in W-types.

  Establishing that Feynman graphs form higher inductive types involves giving
  some uniform description of overlapping divergences, perhaps in terms of
  rewrite systems, and showing that the governing patterns are given by polynomial
  data in a sense compatible with the formalism developed in
  \cite{Lumsdaine-Shulman:semantics-HIT}.  Hopes that such advances are
  achievable stem from the fact that practitioners of quantum field theory have
  already garnered an extremely large body of experience with Feynman graphs, and attention has already been
  given  to the subtleties of overlapping divergences.  I speculate that
  it could be of some importance to sort out these questions.
\end{blanko}

\begin{acknowledgments}
  This contribution is a write-up of my talk at the workshop in Trento on
  {\em Dyson-Schwinger Equations
  in Modern Mathematics \& Physics}.
  I wish to thank the organisers, Mario Pitschmann, Craig Roberts, and Wolfgang
  Lucha, for the wonderful workshop, for the invitation to speak there, and for
  the opportunity to learn a lot about the subject area.  I am grateful to
  Kurusch Ebrahimi-Fard for his patience, over the years, in explaining me some
  rudiments of quantum field theory, and to Nicola Gambino and Egbert Rijke
  for helping me learn type theory.  This work was funded by grant number
  MTM2013-42293-P of Spain.
\end{acknowledgments}

\section*{Appendix: initiality in terms of dependent elimination}

In category theory, an object $I$ of a category $\C$ is called {\em initial}
when, for any other object $A$, there is a unique morphism $I \to A$ in $\C$.

\bigskip

\noindent
{\bf Lemma.}
  {\em Let $\C$ be a category, assumed to admit finite limits.  Then an object
  $I$ of $\C$ is initial if and only if every morphism $C \to I$ has a section.}
\begin{proof}
  Suppose $I$ is initial, and suppose given $p:C \to I$.  Since $I$ is initial,
  there exists a morphism $f:I \to C$.  This morphism $f$ is in fact a 
  section to $p$, because both the composite $p\circ f$ and the identity 
  morphism are morphisms $I \to I$, so by initiality of $I$, they must 
  coincide.
  
  Conversely, suppose every $C \to I$ has a section.  For  an arbitrary
  object $A$, we need to establish that there is precisely one morphism $I \to A$.
  Consider the product $I \times A$; by assumption, this first projection
  $I\times A \to I$ has a section, i.e.~a morphism $I \to I \times A$. 
  Composed with the second projection this gives the existence of $a:I \to
  A$.  To see that it is unique, suppose we have also another, $b:I \to A$.  The
  two morphisms together constitute a morphism $(a,b) : I \to A\times A$.  We wish
  to show that this morphism factors through the diagonal $A \to A \times A$, because
  that is precisely to say that they coincide.  But to find that factorisation
  $$
  \xymatrix{
  & A \ar[d]^{\text{diag}} \\
  I \ar@{..>}[ru] \ar[r]_-{(a,b)} & A\times A
  }$$
  is equivalent to finding a section to the pullback (fibre product)
  $$
  \xymatrix{
  I \times_{A\times A} A \ar[d] \ar[r]& A \ar[d]^{\text{diag}} \\
  I \ar@{..>}@/^1pc/[u] \ar[r]_-{(a,b)} & A\times A  .
  }$$
  But this section exists by assumption.
\end{proof}

\hyphenation{mathe-matisk}


%

\end{document}